# Tuning Optical Orbital Angular Momentum in Optical Superlattice under Electro-optic Effect

**Jie Wang, Jianhong Shi\*, Linghao Tian, Xianfeng Chen\*\***

*Department of Physics; the State Key Laboratory on Fiber Optic Local Area Communication Networks and Advanced Optical Communication Systems,
Shanghai Jiao Tong University. 800, Dong Chuan Road, Shanghai, 200240, China
\*purewater@sjtu.edu.cn, \*\* xfchen@sjtu.edu.cn*

**Abstract:** We introduce a new method to generate and tune the optical orbital angular momentum of a focused Gaussian beam passing through the optical superlattice under the electro-optic effect. The orbital angular momentum (OAM) arises from the curl of polarization in our calculation. We see that adjusting the external electric field, the beam waist radius and the crystal length provides dramatic variation of OAM of light across the transverse section. It is believed that this invention will find its application in optical manipulation area.

## 1. Introduction

Nearly twenty years ago, *Allen et al* [1] discovered a light beam carries both orbital angular momentum (OAM) and spin angular momentum (SAM) for the first time. As a natural character of light, the OAM was considered to arise from the spiral phase distribution of light beam. Since then, it has drawn extensive concerns and interest for its practical applications. The applications range from nonlinear optics [3], optical tweezers [4], and quantum information processing [5], to free space communications [6]. The OAM is not only confined in visible light, but also extended to other wave ranges such as *x*-ray [7], sonic wave [8], and matter wave [9]. Moreover, recent work has shown that the OAM originates from not only the spiral phase distribution, but the curl of polarization of light beam [10, 11, 12] as well. Inducing OAM via spiral phase is easy to achieve by some linear technique, i.e. laser mode converters [13], spiral phase plate [14], computer-generated holograms [15], and Pancharatnam-Berry phase optical elements [16] as well as some nonlinear methods [17]. However, seldom method has been proposed to generate the OAM from the curl of polarization except the original proposal in reference [10], where a pair of orthogonal base vectors is employed to generate radial variant vector fields with hybrid states of polarization carrying optical OAM. Recently, *Tang et al* [18] presents radial variant polarization in transverse section of focused Gaussian light propagating in an OSL under EO effect. In fact, we can use this method to generate and tune OAM.

In this paper, we theoretically study the tuning of the OAM of a focused Gaussian beam after passing through the optical superlattice (e.g. periodically poled lithium niobate (PPLN)) under the electro-optic effect. We found that the OAM distribution on the light transverse section is just relevant to radius. Moreover, OAM can be continuously adjusted through selecting proper applied electric field , waist radius of the Gaussian beam and crystal length.

## 2. Theory and simulation

A monochromatic light field propagating along *x* axis with frequency $\omega$ has a vector potential which can be expressed as $\boldsymbol{A}(y,z) = A(y,z) \times [\alpha(y,z)\hat{e}_y + \beta(y,z)\hat{e}_z]\exp(ikx - i\omega t)$ where the complex light amplitude is expressed by real valued module *u* and phase $\psi$, $\alpha$ and $\beta$ represent states of polarization across the *yz* section of the light field. The momentum flux, under the condition of Lorenz gauge and paraxial approximation, can be described as $\boldsymbol{P} \propto \langle \boldsymbol{E} \times \boldsymbol{H} \rangle$, where $\boldsymbol{E} = i\omega\boldsymbol{A} + i(\omega/k^2)\nabla(\nabla\cdot\boldsymbol{A})$, $\boldsymbol{H} = (\nabla\times\boldsymbol{A})/\mu_0$. As a result, the transverse component of $\boldsymbol{P}$ can be divided into three components as follows

$$\boldsymbol{P}_\perp^{(1)} \propto 2u^2\nabla\psi$$

$$\boldsymbol{P}_\perp^{(2)} \propto iu^2(\alpha\nabla\alpha^* - \alpha^*\nabla\alpha + \beta\nabla\beta^* - \beta^*\nabla\beta)$$

$$\boldsymbol{P}_\perp^{(3)} \propto i\nabla\times[u^2(\alpha\beta^* - \alpha^*\beta)\hat{e}_x]$$

For angular momentum flux, $\boldsymbol{J} \propto \boldsymbol{r}\times\boldsymbol{P}$, accordingly, the *x* component of $\boldsymbol{J}$ is given by

$$J_x^{(1)} \propto 2u^2\partial\psi/\partial\phi \quad (1a)$$

$$J_x^{(2)} \propto iu^2(\alpha\partial\alpha^*/\partial\phi + \beta\partial\beta^*/\partial\phi - c.c.) \quad (1b)$$

$$J_x^{(3)} \propto ir\partial[u^2(\alpha\beta^* - \alpha^*\beta)]/\partial r \qquad (1c)$$

In this work [10], OAM originates from the curl of polarization, which is determined by (1c). In their experiment, a space light modulator is used to add additional phase to generate radial variant polarization carrying OAM. They utilize time-lapse photographs to observe isotropic particles driving the orbiting motion which is the evidence of OAM.

Due to the EO effect, the output beam of focused Gaussian light in PPLN can form spatially radial variant polarization, which was firstly theoretically analyzed by Tang *et al* [18]. Here we use this method to generate and tune OAM. Fig.1 shows the experimental setup of control OAM on the output light section through EO effect by focusing Gaussian beam on the PPLN. An external electric field is applied along the *y*-axis and the monochromatic light beam propagates in the *x* direction. The total field is expressed as follows [20]:

$$\boldsymbol{E}(t) = \boldsymbol{E}(0) + [\boldsymbol{E}(r,x)\exp(-i\omega t) + c.c.]/2 \qquad (2)$$

where $\boldsymbol{E}(0)$ is the dc electric or slow varying electric field. A monochromic light with frequency $\omega$ can be depicted by $[\boldsymbol{E}(r,x)\exp(-i\omega t)/2+c.c.]$, and *c.c.* denotes the complex conjugate. Generally, $\boldsymbol{E}(r,x)$ is considered as composition of two perpendicular vectors:

$$\boldsymbol{E}(r,x) = \boldsymbol{E}_1(r,x)\exp(ik_1 x) + \boldsymbol{E}_2(r,x)\exp(ik_2 x) \qquad (3)$$

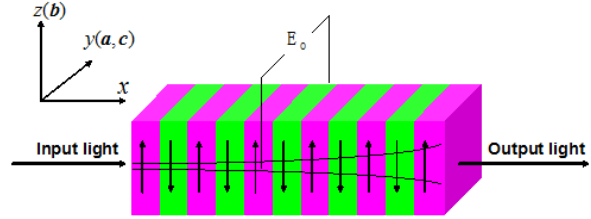

**Fig. 1.** Schematic diagram of generation of OAM of a focused Gaussian beam in an optical superlattice (e.g. a z-cut PPLN) under electro-optic effect. A uniform electric field $\boldsymbol{E}_0$ is applied along the *y*-axis of the PPLN sample. *a* and *b* are unit vectors of two independent electric field components. *c* is the direction of applied electric field. The directions of the polarizations of crystal domains are indicated by arrows.

Gaussian beams can be expressed as $\boldsymbol{E}_j(r,x)=\boldsymbol{B}_j(x)u_j(r,x)$ ($j=1,2$), where $\boldsymbol{B}_j(x)$ are the expansion coefficients of the Laguerre-Gauss modes of zero order, and $u_j(r,x)$ are the Gauss modes [19]. We focused the waist of a Gauss beam on the input surface of the optic superlattice, and the waist radius $W_j$ of $\boldsymbol{E}_j$ are equal ($W_j=W_0$). So the modes are given by

$$u_j(r,x) = \sqrt{\frac{2}{\pi}}\frac{1}{W_0(1-i(2x/b_j))}\exp\{-\frac{r^2}{W_0^2[1-i(2x/b_j)]}\} \qquad (4a)$$

$$\boldsymbol{B}_1(x) = \sqrt{\omega/n_1}\,g_1(x)\boldsymbol{a} \quad \boldsymbol{B}_2(x) = \sqrt{\omega/n_2}\,g_2(x)\boldsymbol{b} \quad \boldsymbol{E}(0) = E_0\boldsymbol{c} \qquad (4b)$$

where the confocal parameters $b_j=k_j W_0^2$; $\boldsymbol{a}$, $\boldsymbol{b}$ and $\boldsymbol{c}$ are three unit vectors and $\boldsymbol{a}\cdot\boldsymbol{b}=0$; $g_1(x)$ and $g_2(x)$ are the normalized amplitudes of the two independent wave components. When paraxial approximation holds, similarly to reference [18], the normalized amplitudes satisfy coupling equations as follows

$$\frac{dg_1(x)}{dz} = -id_1 g_2(x)f_1\frac{1}{1+(x/b_1)^2(1-n_1/n_2)^2} - id_2 f(x)g_1(x)$$

$$\frac{dg_2(x)}{dx} = -id_3 g_1(x)f_1\frac{1}{1+(x/b_1)^2(1-n_1/n_2)^2} - id_4 f(x)g_2(x) \qquad (5)$$

with $d_1 = \frac{k_0}{2\sqrt{n_1 n_2}} \gamma_{eff1} E_0$, $d_2 = \frac{k_0}{2n_1} \gamma_{eff2} E_0$, $d_3 = d_1$, $d_4 = \frac{k_0}{2n_2} \gamma_{eff3} E_0$,

$$f_1 = \frac{1}{L}\int_0^L f(x)\exp[iRx + \varphi(x)]dx \quad \varphi(x) = \arg\{[1 \pm i(x/b_1)(1 - n_1/n_2)]^{-1}\},$$

$$f(x) = \text{sgn}(\text{Re}\{[1 + i(x/b_1)(1 - n_1/n_2)]^{-1}\exp(i\Delta kx)\}), \text{ and } \Delta k = k_2 - k_1$$

where $f(x)$ is the structure function, and it equals 1 or -1 when $x$ falls in the positive or negative domain, and $f_1$ is the Fourier coefficient; $L$ is the length of crystal; $R$ is the reciprocal vector provided by the optical superlattice; $\gamma_{effi}(i=1,2,3)$ is effective electro-optic coefficients as those in [20]. When $x \ll b_j$, the coupling equations will degenerate to plane wave approximation.

For a Gaussian beam, if we set the beam waist at (0,0,0), its vector potential $A$ under the Davis procedure is taken to be the form [21]

$$A(r) = (iE/k)\psi(x,y,z)\exp(ikx)$$

Where $E$ is the electric field, the slow varying function $\psi$ obeys from the equation $\nabla^2 \psi + 2ik\partial \psi / \partial z = 0$, and its solution is

$$\psi(r) = \psi_0(r) + s^2 \psi_2(r) + \cdots \qquad (6)$$

where $s=W_0/b=1/(kW_0)$ is the beam parameter equal to the waist radius $W_0$ divided by the confocal parameter $b$. The lowest order functions are given by $\psi_0 = iQ\exp(-iQ\rho^2)$ and $\psi_2 = (2iQ + i\rho^4 Q^3)\psi_0$, where $\rho^2 = \xi^2 + \eta^2$, $Q = (i - 2\zeta)^{-1}$, $\xi=y/W_0$, $\eta=z/W_0$, and $\zeta=x/b=sx/W_0$.

In our scheme, we obtain the two vectors potential

$$A_1(r) = \frac{iE_1}{k_1} \frac{1}{1 + i\frac{2x}{k_1 W_0^2}} \exp\left\{-\frac{r^2}{W_0^2(1 + i\frac{2x}{k_1 W_0^2})}\right\}\exp(ik_1 x) \qquad (7a)$$

$$A_2(r) = \frac{iE_2}{k_2} \frac{1}{1 + i\frac{2x}{k_2 W_0^2}} \exp\left\{-\frac{r^2}{W_0^2(1 + i\frac{2x}{k_2 W_0^2})}\right\}\exp(ik_2 x) \qquad (7b)$$

this result is obtained by neglecting the second order $\psi_2$ when $s<0.005$.

For equation (1) and (7), there is no azimuth angle factor $\phi$ included in our vector potential, so both $J_x^{(1)}$ and $J_x^{(2)}$ are zero. The OAM is determined by (1c), which is merely related to the radial distributions of the states of polarization. Inserting equation (7) into (1c), we arrive at

$$J_x^{(3)} \propto \frac{2i}{W_0^2}\left|\frac{E_1}{k_1}C_1\right|\left|\frac{E_2}{k_2}C_2\right|r^2\{D_1 \exp(-\frac{r^2}{W_0^2}D_1) - D_2 \exp(-\frac{r^2}{W_0^2}D_2)\} \qquad (8)$$

where $C_j=(1+i2x/b_j)^{-1}$ ($j=1,2$), $D_1=C_1+C_2^*$, $D_2=D_1^*$.

From equation (8), we find the OAM varies with change of external electric field, beam waist and length of crystal.

In the following, without loss of generality, we take 2.5*cm* PPLN, with the period fixed as an example to analyze the above expression. The wavelength of the light is fixed at 632.8*nm*, and the refractive indexes $n_1$=2.2884 and $n_2$=2.2019 are calculated from Sellmeier equations for pure LiNbO$_3$ [22] at room temperature 298K. To illustrate the influence of external electric field on the OAM of light beam, we take the beam waist as 15*μm*, and set the initial condition of incident light as $A_1(0)=0$, $A_2(0)=1$, with the extraordinary light incidence. The coupling between ordinary and extraordinary light is controlled by the external electric field. When the extraordinary light is launched into the PPLN, the amplitude of the ordinary light coupled out from extraordinary light, as shown in Fig. 2(a). We can see that, when the external electric field rises to 74$V$/mm, the extraordinary wave can be fully transferred to ordinary light. An interesting phenomenon is found in Fig. 2(b): (1) OAM is 0 no matter how the applied electric field is when $r$ =0 and $r$ =105$\mu m$ because eq.(8) is zero. (2) The distribution of OAM appearances do not change, but OAM varies under different applied electric fields $E_0$. Note that Fig. 3 shows the OAM distribution of the light under different applied electric field, and we find it varies periodically with the increment of external electric field. At the output surface of the crystal, the beam radius is about 150$\mu m$. For every figure $J$ always reach the maxium at the external ring. As shown in Fig.3 (a) and (f), the OAM of light is zero across the beam section because the output light is linearly polarized, corresponding to the external electric field $E_0$=0$V$/$mm$ and 74$V$/$mm$, respectively. In the first condition, the output beam remains its incident polarization without the effect of external electric field, while extraordinary light converts to ordinary wave completely in the second condition. Fig. 3(b-e) show OAM distribution when $E_0$ takes different voltage. The maximum $J$ is reached when the intensity of extraordinary wave equals to ordinary wave at $E_0$=37$V$/$mm$ in Fig.3(c).

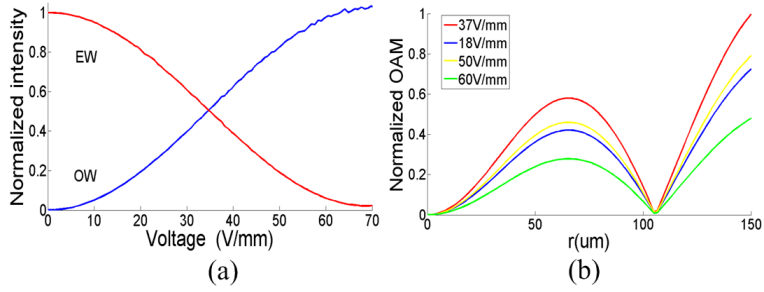

**Fig. 2.** (a)The normalized extraordinary (EW) and ordinary wave (OW) intensity of output beam vs external electric field. The red and blue line represents EW and OW respectively.(b)The normalized OAM of output light beam passing through the PPLN controlled by different external electric field.

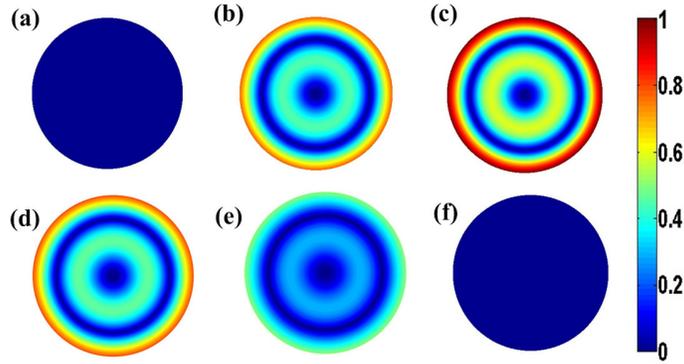

**Fig. 3.** The spatial distribution of OAM of output beam for different $E_0$ with λ=632.8nm, T=298$K$,L=2.5$cm$, $W_0$=15$\mu m$, $r_{out}$=150$\mu m$; (a)$E_0$=0$V$/$mm$;(b) $E_0$=18$V$/$mm$; (c) $E_0$=37$V$/$mm$; (d) $E_0$=50$V$/$mm$; (e)$E_0$=60$V$/$mm$;(f) $E_0$=74$V$/$mm$.

Not only the applied electric field, but the beam waist also influences OAM of output beam. We assume extraordinary light incidence and fix $E_0$=37$V$/$mm$, $L$=2.5$cm$. The OAM distributions are shown in Fig. 4(a-d) with $W_0$ taken as 15$\mu m$, 30$\mu m$, 60$\mu m$, 120$\mu m$ and radius of output beams change accordingly. The maximum OAM becomes smaller with increment of $W_0$ and will decrease to 0 when the waist is infinite for the focused Gaussian beams will degenerate to plane wave. Hence we achieve spatial inhomogeneity of polarization controlled by external electric field, through which the continuous change of OAM distribution can be manipulated.

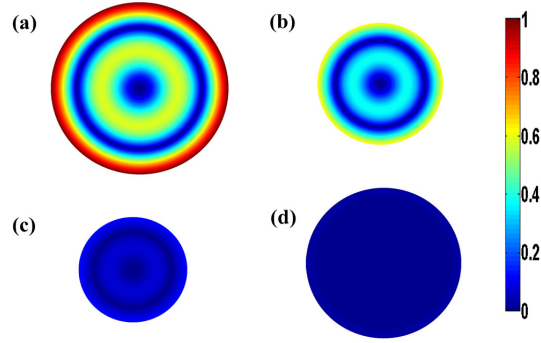

**Fig.4.** The spatial distribution of OAM of output beam for different $W_0$ with $\lambda$=632.8nm, T=298K, $L$=2.5cm, and $E_0$=37$V/mm$ fixed.(a)$W_0$=15$um$,$r_{out}$=150$\mu m$;(b)$W_0$=30$um$,$r_{out}$=80$\mu m$;(c)$W_0$=60$um$,$r_{out}$=70$\mu m$;(d) $W_0$=120$um$, $r_{out}$=120$\mu m$.

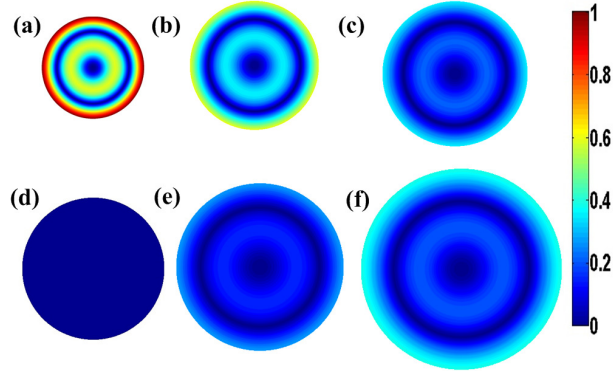

**Fig. 5.** The spatial distribution of OAM of output beam for different crystal lengths $L$ with $\lambda$=632.8$nm$,$T$=298$K$,$W_0$=15$\mu m$,$E_0$=37$V/mm$.(a)$L$=2.5$cm$,$r_{out}$=150$\mu m$;(b)$L$=3.4$cm$,$r_{out}$=200$\mu m$;(c) $L$=4.0$cm$,$r_{out}$=235$\mu m$;(d)$L$=5.0$cm$,$r_{out}$=290$\mu m$;(e)$L$=5.8$cm$, $r_{out}$=335$\mu m$; (e)$L$=6.6$cm$, $r_{out}$=400$\mu m$;

Apart from voltage and beam waist, we investigate the OAM by selecting different lengths of crystal under the condition that the applied electric field is 37$V/mm$ and beam waist radius is 15$\mu m$. When the crystal length becomes longer, the radius of the output beam becomes larger. The maximum OAM in Fig. 5 (a-f) is 1, 0.71, 0.35, 0, 0.27, and 0.41, corresponding to discrete crystal lengths, respectively. In addition, high OAM area is concentrated in the external ring. Especially, if $L$ takes 5.0$cm$, the e polarization is modulated to o light wave absolutely. Thus the OAM is 0.

### 3. Conclusions

In conclusion, we systematically investigate the OAM behavior of a focused Gaussian beam after passing the optical superlattice under the electro-optic effect. This work proposes a new method to generate and manipulate the OAM of light. The OAM can be tuned continuously by changing the applied electric field consecutively. Other parameters, i.e. beam waist and crystal length are also proved to be effective means to control the OAM of light beam. We believe that this method will find its splendid application as optical tweezers to manipulate the micro particles.

### 4. Acknowledgements

This research was supported by the National Natural Science Foundation of China and NSAF (Contract No. 10876019), the Shanghai Leading Academic Discipline Project (No. B201)